# High temperature magnetic stabilization of cobalt nanoparticles by an antiferromagnetic proximity effect

José A. De Toro,[1,*] Daniel P. Marques[1], Pablo Muñiz[1], Vassil Skumryev,[2,3] Jordi Sort,[2,3] Dominique Givord,[4,5,6,*] Josep Nogués[3,7,*]

[1]Instituto Regional de Investigación Científica Aplicada (IRICA) and Departamento de Física Aplicada, Universidad de Castilla-La Mancha, E-13071 Ciudad Real, Spain
[2]Departament de Física, Universitat Autònoma de Barcelona, E-08193 Bellaterra (Barcelona), Spain
[3]Institució Catalana de Recerca i Estudis Avançats (ICREA), Barcelona, Spain
[4]Univ. Grenoble Alpes, Institut NEEL, F-38042 Grenoble, France
[5] CNRS, Institut NEEL, F-38042 Grenoble, France
[6]Instituto de Fisica, Universidade Federal do Rio de Janeiro, Rio de Janeiro RJ, Brasil
[7]ICN2 – Institut Catala de Nanociencia i Nanotecnologia, Campus UAB, E-08193 Bellaterra (Barcelona), Spain

**Abstract**

Thermal activation tends to destroy the magnetic stability of small magnetic nanoparticles, with crucial implications in ultra-high density recording among other applications. Here we demonstrate that low blocking temperature ferromagnetic (FM) Co nanoparticles ($T_B$<70 K) become magnetically stable above 400 K when embedded in a high Néel temperature antiferromagnetic (AFM) NiO matrix. The origin of this remarkable $T_B$ enhancement is due to a magnetic proximity effect between a thin CoO shell (with low Néel temperature, $T_N$; and high anisotropy, $K_{AFM}$) surrounding the Co nanoparticles and the NiO matrix (with high $T_N$ but low $K_{AFM}$). This proximity effect yields an effective AFM with an apparent $T_N$ beyond that of bulk CoO, and an enhanced anisotropy compared to NiO. In turn, the Co core FM moment is stabilized against thermal fluctuations via core-shell exchange-bias coupling, leading to the observed $T_B$ increase. Mean-field calculations provide a semi-quantitative understanding of this magnetic- proximity stabilization mechanism.

The current miniaturization trend in magnetic applications has led to a quest to suppress spontaneous thermal fluctuations (superparamagnetism) in ever-smaller nanostructures [1-5], which is a clear example of the fundamental efforts of condensed matter physics to meet technological challenges [6] (e.g., the continued growth of recording density [7]). Despite the foreseeable change of recording paradigm from continuous to patterned media, where each bit is recorded in an individual nanostructure [7], the key for sustained storage density increase will remain the introduction of progressively more anisotropic (high K) materials [8], which allow for magnetic stability at very small volumes, V (i.e., blocking temperature, $T_B \propto KV$, above room temperature, RT). Two main strategies are largely investigated to achieve high K (both of them with implications in other active technologies beyond information storage, such as permanent magnets, magnetic hyperthermia or even sensors [5,9-11]): (i) the use of compounds with intrinsically high magnetocrystalline anisotropy (such as FePt [3,8]) and (ii) the design of exchange-coupled nanocomposites [4,12-29]. Unfortunately, most high-K materials require high-temperature annealing processes to obtain the desired phase, which could hamper their implementation in certain structures. Thus, FM-AFM exchange coupling alternatives may be an appealing option. In fact, it has been demonstrated [4] that ferromagnetic-antiferromagnetic (FM-AFM) interfacial exchange-coupling is an effective method, later patented by Seagate [12], to increase the effective K of FM nanoparticles. However, a $T_B$ enhancement beyond RT using this approach has been rarely reported [22-26] (where often broad particle size distribution can partly account for the "apparent" $T_B$ increase [22-25]). The reason for this scarcity is that high Néel temperature ($T_N$) AFMs tend to have a low anisotropy constant (e.g., NiO), and vice versa (e.g., CoO), while substantial values of *both* properties are required for high-temperature stabilization. This limitation could, in principle, be overcome by exploiting proximity effects, i.e., the interfacial synergetic hybridization of the properties of two AFM materials having complementary properties (here, high $T_N$ and high K). Although this phenomenon is best known in superconductivity [30], proximity effects in bi- or multi-layered magnetic systems (i.e., *magnetic* proximity effects) have also been studied [31]. In contrast, and despite their strong technological presence, proximity effects involving nanoparticles have been hardly explored [32,33].

In this Letter we demonstrate a proximity effect between two AFMs (a CoO shell and a NiO matrix) on FM particles (Co) and the resulting thermal stabilization of the NPs well above RT (with an ~10-fold enhancement of $T_B$ to exceed 400 K), and propose a mean-field model to gain insight into the nature of such AFM proximity effect.

Three films of Co/CoO core/shell nanoparticles [4,34-36] (5-7 nm) highly dispersed in an AFM NiO matrix (S-series) –Figure 1(a)– (or in a Nb matrix, for reference, R-series) were grown by combining inert gas condensation (Co nanoparticles) and RF-sputtering (NiO and Nb) [34-40]. The digits in the sample names refer to the cluster source power (W), which, together with the occasional use of a carrier gas (He), was varied to control the nanoparticle size [35].

The low-temperature hysteresis loops of the Co/CoO-NiO samples (S-series) measured after field cooling are shown in Figure 1(b). The loops show rather large coercivities ($\mu_0 H_C \sim 0.4$ T) and loop shifts (i.e, $H_E$, exchange bias) $\mu_0 H_E \sim 0.4$ T. In contrast, the loops exhibit a rather small vertical shift (less than 1% of $M_S$). In the reference samples, where NiO is replaced by Nb (R-series), $H_C$ is considerably smaller ($\mu_0 H_C \sim 10$ mT) and no loop shifts are observed [Figure 1(b) inset].

Remarkably, the T=300 K hysteresis loops shown in Figure 1(c) evidence that the samples are not superparamagnetic (i.e., with remanence, $M_R$, and $H_C>0$). Not only is $\mu_0 H_C \sim 6$mT, but $H_E$ is surprisingly large (e.g., $\mu_0 H_E = 14$ mT for *S50He*). In contrast, the reference samples have vanishing $M_R$ and $H_C$ at T = 300 K, revealing their superparamagnetic state (Figure S1 [34]).

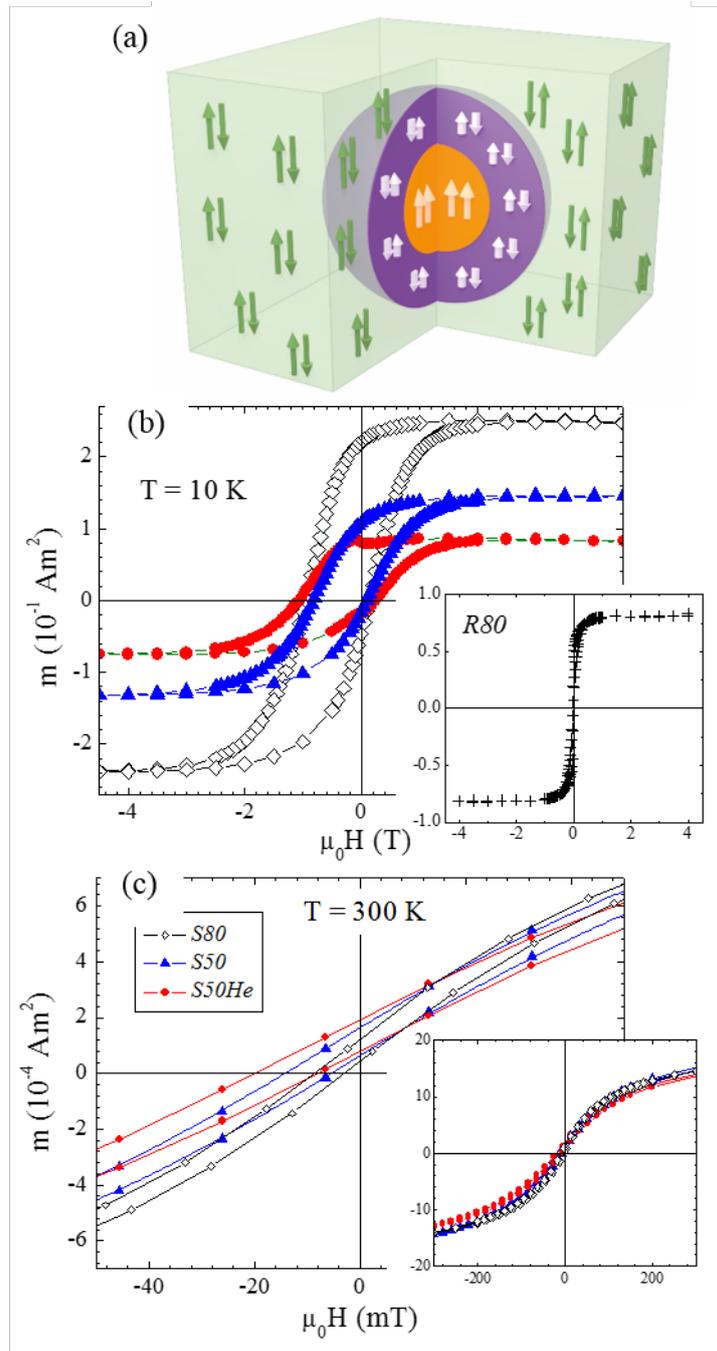

**Figure 1**. (a) Schematic representation of the Co/CoO-NiO sample. (b) Field-cooled hysteresis loops of the Co/CoO-NiO samples (S-series) at 10 K. Shown in the inset is the 10 K hysteresis loop of R80. (c) Field-cooled hysteresis loops of the same samples at 300 K. Shown in the inset are the same loops up to higher fields.

To assess the effect of coupling on the value of the superparamagnetic $T_B$, the FC/ZFC temperature dependence of the magnetization, M(T), was measured for two Co/CoO-NiO and Co/CoO-Nb films (*S80* and *R80* -largest particles- and *S50He* and *R50He* -smallest particles-) [Figure 2(a,b)]. The M(T) curves of the reference samples show the typical behaviour of nanometric Co nanoparticles: $T_B$ (taken as the maximum of the ZFC curve) is low and increases with particle size [i.e., $T_B$(*R50He*)≈35 K and $T_B$(*R80*)≈70 K]. In contrast, $T_B$ for the Co/CoO-NiO samples is beyond RT [$T_B$(*S80*)≈360 K] and even above 400 K (the maximum experimentally attainable temperature) for the case of *S50He*. The temperature dependence of $H_E$ [Figure 2(c)], which establishes the exchange bias blocking temperature, $T_B[H_E]$, shows a similar trend to the superparamagnetic $T_B$, with $H_E$ remaining finite probably above T=400 K.

The present results demonstrate that Co nanoparticles of a few nm can be made magnetically stable above RT, up to at least T=400 K [41]. The origin of the enhanced magnetic stability must reside in some coupling existing between the Co nanoparticles and the high-$T_N$ AFM matrix, NiO ($T_N$=520 K), since using CoO alone as matrix limits the $T_B$ enhancement to 290 K [$T_N$(CoO)] [4]. However, NiO is known to have a low anisotropy [42], leading to small $H_E$ and low $T_B[H_E]$ (often below RT) [25,27,43-46]. This highlights that using a high-$T_N$ material is not sufficient in itself to reach high-temperature stability.

The first indication of the origin of the observed effects is the very large $H_E$ measured in the Co/CoO-NiO series at T=10 K. NiO alone cannot induce such high $H_E$ values, hence, the highly anisotropic CoO shell must be involved in the $H_E$ enhancement. However, isolated Co/CoO nanoparticles with a thin (natural oxidation) CoO shell usually exhibit very small $H_E$ [39,40]. Three main types of processes have been proposed to achieve large $H_E$ in Co/CoO systems [4,13-15,47]: (i) forced oxidation of the Co particles to form thick AFM CoO shells [13-15], (ii) matching the crystallographic structure between the CoO shell and the matrix (which structurally stabilizes the CoO shell) [47] and (iii) coupling the CoO shell to an AFM matrix (magnetic stabilization) [4]. Our low-oxygen synthesis method allows to safely rule out the first possibility [13,15]. Since the crystalline structures of NiO and CoO are similar,

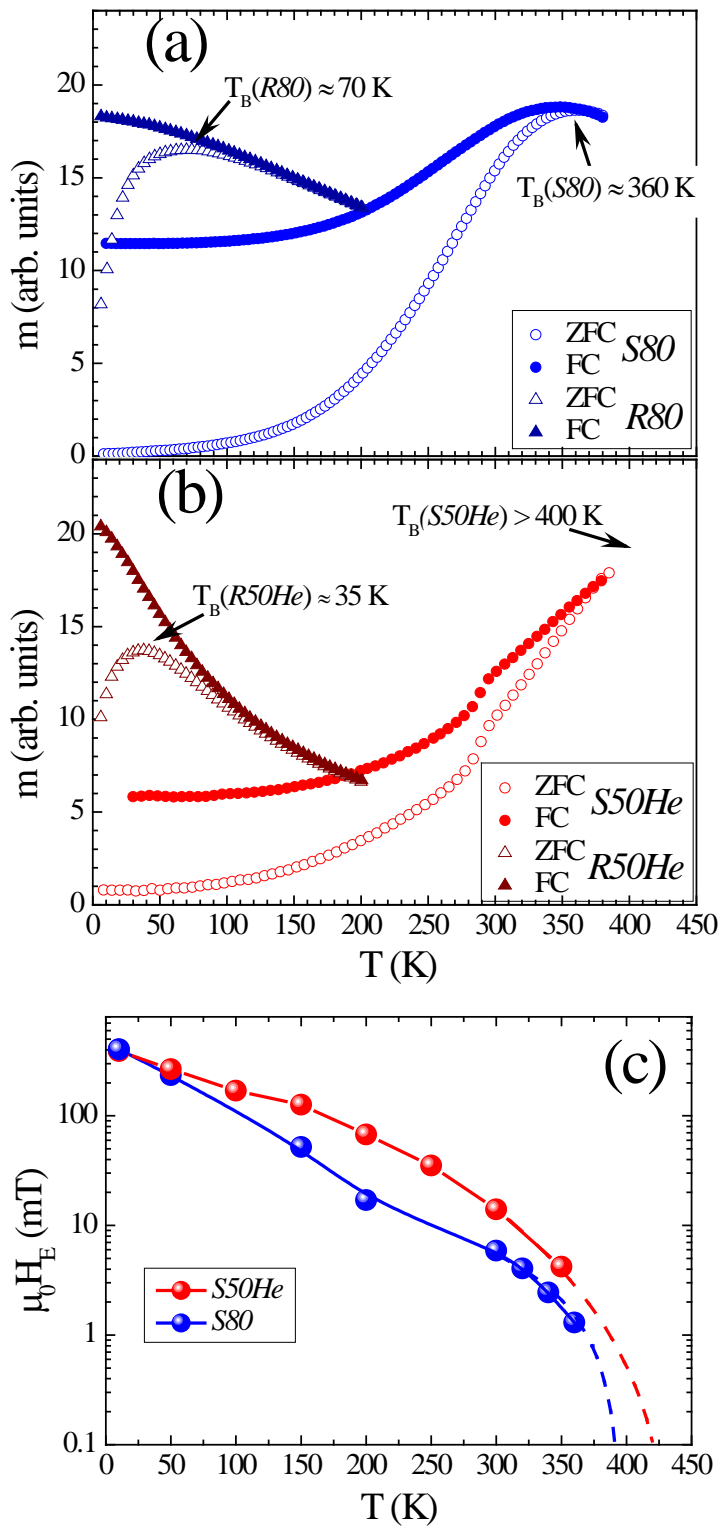

Figure 2. FC/ZFC magnetization curves measured in 20 mT for (a) *S80* and *R80* and (b) *S50He* and *R50He*. The $T_B$ values of the four samples are highlighted by arrows. (c)

Temperature dependence of $H_E$ for *S80* and *S50He*. The continous lines are guides to the eye. The dashed curves are tentative extrapolations of $H_E(T)$ to hint $T_B[H_E]$.

both structural and magnetic stabilizations of the CoO shell are a priori plausible. However, the high values (above 400 K) of the superparamagnetic $T_B$ and $T_B[H_E]$ imply that the magnetic stabilization cannot be solely a structural effect. Indeed, NiO may structurally stabilize CoO, nevertheless the $T_N$ of CoO ($T_N$=290 K) is exceedingly low to cause the observed high-temperature effects. Consequently, the outstanding enhancement of $T_B$ of the Co nanoparticles must be a combined magnetic effect involving both the CoO shell and the NiO matrix.

In thin film systems it has been previously observed that $H_E$ and $T_B[H_E]$ of NiFe/NiO bilayers can be tailored by inserting a thin CoO layer at the interface between both layers, i.e., NiFe/CoO/NiO [48,49]. When the CoO interfacial layer remains below 3 nm, $T_B[H_E]$ persists above T=400 K, whereas for thicker CoO, it drops quickly to $T_N$(CoO). This effect can be understood as a *magnetic proximity effect* [31], where the overall properties of $AFM_1/AFM_2$ systems are the combination of both counterparts [50-52]. This concept has been applied recently to other types of AFMs such as IrMn/FeMn [53] and it must take place in the Co/CoO-NiO system, where the overall $T_B$ is determined by the combined effect of the CoO shell coupled to the NiO matrix. However, to explain the high-temperature stability of the Co nanoparticles, a polarization of the Co AFM moments in the CoO shell is not sufficient; the overall anisotropy of the CoO-NiO couple, ultimately felt by the Co particles, must also remain sufficiently high. Consequently, the proximity effect between CoO and NiO has a two-fold consequence where both the Co induced magnetization and the overall anisotropy are involved [54].

For systems composed of FM nanoparticles embedded in an AFM matrix, $H_E$ is classically expressed as:

$$\mu_0 H_E M_{FM} V = \gamma A \qquad \text{(Eq. 1)}$$

where $M_{FM}$ is the FM magnetization, $V$ is the volume of the ferromagnet, $\gamma$ is the interfacial coupling energy per unit surface area, and $A$ the associated surface area. The evaluation of $\gamma_0$, the 0 K coupling energy, constitutes the major difficulty in the analysis of exchange-bias systems. Here, $\gamma_0$ is only taken as an experimental parameter. Naively, the temperature dependence of $\gamma$ should be proportional to the interfacial AFM staggered magnetization (neglecting the temperature dependence of the FM Co nanoparticle magnetization). Given the complexity of experimentally obtaining the surface magnetization of the CoO nanoparticles, we have developed a simple molecular field model [34]. The mean field was determined by considering exchange-interactions between first-neighbours, in agreement with the short-range nature of super-exchange interactions. An excellent agreement was obtained between the calculated temperature dependence of the CoO bulk staggered magnetization, assuming S=3/2, and previous experimental results [55] (Figure S2a [34]). The temperature dependence of the surface magnetization was then calculated by assuming that, for surface atoms, the number of neighbours is reduced from 12 in the bulk to 9. The temperature dependence of the surface magnetization (Figure S2b [34]) is reminiscent of the temperature dependence of the remanent magnetization in CoO nanoparticles, which has been related to surface magnetic moments [55]. Additionally, the calculated variation of the surface magnetization reproduces correctly the temperature dependence of $H_E$ in the Co/CoO-CoO system (i.e., Co/CoO nanoparticles embedded in a CoO matrix [4]) in the whole temperature range (compare the calculated temperature dependence of the CoO surface magnetization in Figure S2 [34] to the experimental $\mu_0 H_E(T)$ in Figure S3 [34]).

Obviously, expression (1) cannot explain the sizable $H_E$ measured in the Co/CoO-NiO system at temperatures above the $T_N$ of CoO, if the CoO shell has the same properties as in Co/CoO (Figure 2c). Considering that $T_N$ of NiO (520 K) is much higher than that of CoO, it is natural to attribute the persistence of exchange bias effects to a polarization of the Co moments by the Ni ones. To describe such magnetic proximity effect, the molecular field model was applied to a stack of atomic shells covering a Co sphere of 5 nm diameter. The 5 most external shells are assumed made of pure NiO, they are followed by 3 intermixed shells where the fraction of Ni atoms decreases from 0.75 to 0.5 and 0.25, and by 2 shells of pure CoO. The assumed CoO/NiO interlayer mixing is consistent with reported observations in CoO/NiO multilayers prepared by sputtering [52]. Moreover, the total pure CoO equivalent thickness (≈1 nm corresponding to 2 pure CoO layers, each 0.268 nm thick, + 1.5 equivalent CoO layers from the 3 intermixed layers) is consistent with the oxygen-poor synthesis conditions of the nanoparticles. In the model, a given atom has 12 neighbours in total: 6 neighbours in the shell it belongs to, and $3\alpha_p$ ($\alpha_m$) atoms in the preceding (next) shells, where the coefficients $\alpha_p$ ($\alpha_m$) are proportional to the respective surface area of each considered shell [34]. Calculations then revealed that a significant magnetization is maintained in CoO above its bulk $T_N$ (red line in Figure S2b [34]), via the proximity effect with NiO. Since the Co atoms in CoO at the interface with the core are directly exchange-coupled to the Co FM core (consequently, directly involved in exchange-bias), the existence of a significant CoO staggered magnetization above $T_N(CoO)$ directly accounts for the persistence of exchange-bias in this temperature range. Note that due to the short-range nature of the interactions, the NiO-induced polarization of CoO at the CoO/Co interface becomes negligible if more than two non-intermixed CoO layers are considered.

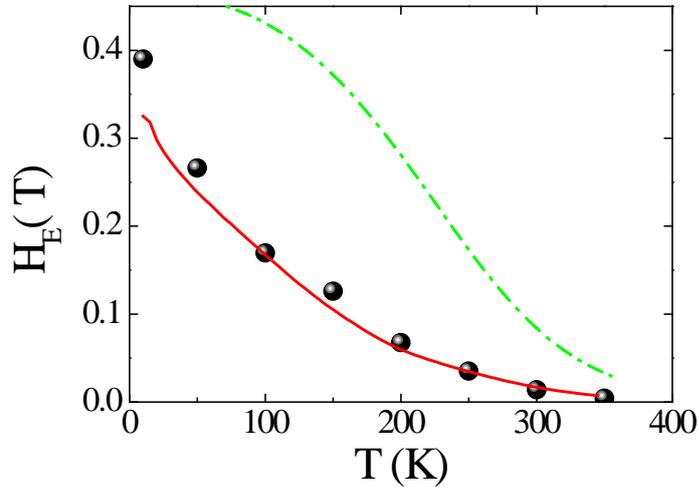

**Figure 3**. Temperature dependence of $H_E$ for Co/CoO-NiO: Experimental data for sample S50He (●); Calculated temperature dependence of $H_E$, neglecting thermal activation (— · —) and taking thermal activation into account (———).

Although this calculation demonstrates that proximity effects in Co/CoO-NiO can account for $H_E$ above $T_N(CoO)$, it does not explain the rapid decrease of $H_E$ with increasing temperature in Co/CoO-NiO compared to Co/CoO-CoO (compare Figure 3 and Figure S3 [34]). Actually, the AFM component in exchange biased systems is usually composed of nanosized grains, which are prone to superparamagnetic effects [56]. To account for the possible existence of superparamagnetic CoO grains, a reduction coefficient must be applied to $H_E$ derived from Equation (1), which does not include thermal activation effects:

$$H_E = H_E^0 f^{2/3} \qquad (Eq.\ 2),$$

where the parameter f represents the volume fraction of AFM grains involved in FM-AFM coupling and $H_E^0$ represents $H_E$ neglecting thermal activation effects. The 2/3 power accounts for the interfacial nature of the FM-AFM coupling. The condition for superparamagnetism is taken as $\Delta E < 25 k_B T$, where $\Delta E$ is the energy barrier and $k_B$ the Boltzmann constant. The main term in $\Delta E$ is the anisotropy energy $K_{AFM} V_{AFM}$, where $K_{AFM}$ and $V_{AFM}$ are the anisotropy and volume of the AFM grains. The AFM grains are assumed composed of a CoO-NiO mixture, the minor CoO fraction being at the interface with the FM nanoparticle. The NiO magnetocrystalline anisotropy is considered negligible [42], thus $K_{AFM} V_{AFM} \approx K_{CoO} V_{CoO}$, where $K_{CoO}$ is the CoO magnetocrystalline anisotropy and $V_{CoO}$ is the volume of the CoO part in each of the oxide grains forming the shell [34]. A Gaussian distribution of energy barriers is assumed. The best fit to the experimental data was obtained for $K_{CoO} V_{CoO} = 0.5\ 10^{-20}$ J, with a standard deviation $\sigma = 0.25\ 10^{-20}$ J, at 0 K. Assuming the bulk CoO magnetocrystalline anisotropy value at low temperatures, $K_{CoO} = 0.4 \, 10^7$ J/m$^3$ [42], yields $V_{CoO} = 1.25$ nm$^3$, which, in turn, corresponds to a CoO layer approximately 0.25 nm thick for cylindrical nanograins 2.5 nm in diameter [34]. This value compares reasonably with the 1 nm CoO equivalent thickness assumed in the calculation of proximity effects, given the uncertainty in the number of AFM grains forming the shell and the reduced magnetocrystalline anisotropy typically found in CoO thin shells due to reduced crystallinity [39,40] (which would actually imply thicker CoO grains). The temperature dependence of the anisotropy of the AFM, $K_{AFM}$, is assumed proportional to the cube of the reduced staggered magnetization as expected for 2$^{nd}$ order anisotropy. The experimental $H_E(T)$ data is semi-quantitatively reproduced (Figure 3) assuming 5 nm diameter Co nanoparticles in agreement with TEM observations (Fig. S4) and $\gamma_0 = 1.1 \times 10^{-3}$ J/m$^2$, consistent with literature data for Co/CoO [57]. The success of the model demonstrates that the rapid decreases of $H_E$ with increasing temperature is linked to the moderate average anisotropy energy of the effective (mixed CoO-NiO) AFM grains, exchange-coupled to the Co nanoparticles. Thermal activation effects in such AFM grains (yielding a distribution of $T_B[H_E]$) are also evidenced by the shape of the ZFC M(T) curves (Figure 2), where the magnetization increases smoothly from low temperatures, in contrast with the relatively abrupt increase (unblocking) observed in exchange-biased systems where a *single* $T_B[H_E]$ value is expected [58,59].

As a consistency test of our model, the temperature dependence of $H_E$ in the Co/CoO-CoO system was re-calculated using the same parameters as above (red line in Figure S3 [34]). As in Co/CoO-NiO, the calculated curves (with the interfacial coupling coefficient $\gamma_0$ as an adjustable parameter) give fair account of the experimental data. The theoretical curves obtained with $\gamma_0 = 1.3 \times 10^{-3}$ J/m$^2$, respectively accounting and neglecting thermal activation, differ only slightly close to $T_N$ (Figure S2 [34]), indicating that in the Co/CoO-CoO system, only a minor fraction of the AFM grains become superparamagnetic as temperature is increased. This situation is related to the high $K_{AFM}$ characteristic of CoO [42]. Altogether, these results reflect the dual role of CoO and NiO in the magnetic stabilization of Co nanoparticles, i.e., while NiO contributes with high-$T_N$, CoO supplies the high anisotropy.

In conclusion, we have presented the foremost example of exchange-bias particle stabilization exploiting magnetic proximity effects. Co/CoO core/shell (~5-7 nm) nanoparticles with blocking temperatures below 70 K have been stabilized well beyond 400 K by combining high-anisotropy CoO and high-$T_N$ NiO antiferromagnets in a shell-matrix configuration which provides an AFM anisotropy at the interface strong enough to enhance the effective anisotropy of the Co cores. A mean-field model, corrected for thermal activation effects, closely reproduces the experimental exchange-bias data, corroborating the above interpretation and illustrating the nature of the proposed proximity effect. The results presented in this study constitute a striking illustration of how a subtle combination of interactions may permit the occurrence of unique magnetic properties by exploiting proximity effects in magnetism. A similar approach could be applied to other composite systems, in and beyond magnetism, where proximity effects may be engineered to enhance material's functionality [30,60-64].

**Acknowledgements**

JADT thanks P.S. Normile and R. López Antón for useful discussion and J.A. González and J.P. Andrés for technical advice. This work has been supported by projects from the Junta de Comunidades de Castilla-La Mancha [PEII11-0226-8769] and from the Generalitat de Catalunya (2014-SGR-1015). D.P. Marques acknowledges the support from the Brazilian CNPQ. ICN2 acknowledges support from the Severo Ochoa Program (MINECO, Grant SEV-2013-0295).

**References**

*****Electronic mail:** JoseAngel.Toro@uclm.es; Dominique.Givord@neel.cnrs.fr; Josep.Nogues@uab.cat

# Supplemental Material

**Experimental Details**

*Deposition Conditions*. Films about 350 nm thick, were grown by high-speed sequential deposition of inert gas-condensed Co nanoparticles (in a modified commercial cluster-source) and a rf-sputtered NiO matrix (from a NiO target) on to thermally oxidized Si(100) substrates using a rotating (0.3 Hz) sample holder [1,2]. Two of the samples were grown using different sputtering powers in the cluster source (50 W – *S50*; 80 W – *S80*). Note that increasing the sputtering power increases the average particle size [3] and the deposition rate from 0.2 (50 W) to 0.4 Å/s (80 W). The third sample was grown at 50 W but using twice as much carrier gas (He) – *S50He*– as in the other samples (5 sccm) with a view of obtaining smaller nanoparticle size [3]. Importantly, oxygen is partially released during the deposition of oxide materials (NiO in our case) by plasma techniques [4] so that the Co nanoparticles partially oxidize to form Co/CoO core/shell nanoparticles. This leads to a core-shell structure of Co/CoO nanoparticles embedded in a NiO matrix (Co/CoO-NiO). Note that in the present conditions the shell grows as CoO and not $Co_3O_4$ [1,2]. Reference samples (named analogously, but starting by *R*) using a niobium matrix instead of NiO (while keeping the same cluster-source parameters for the nanoparticle synthesis) were also prepared. The NiO and Nb matrix targets were sputtered at 150 and 100 W, respectively, which provides deposition rates much larger than those of the Co nanoparticles.

The Co/Ni composition ratio, as derived from energy dispersive microanalysis, was lower than 5%, which implies that the concentration of Co nanoparticles is sufficiently dilute to safely neglect interparticle interactions as well as exchange-bias *connectivity* effects [5,6] resulting from direct contact between the nanoparticles.

*Morphological characterization.* Transmission electron microscopy (TEM) (FEI Tecnai F20 S/TEM operating at 200 kV) was used to estimate the particle size using grids exposed to the nanoparticle beam after the film deposition. The TEM characterization of the samples shows that the particle diameter ranges from ~5 nm (*S50He*) to ~7 nm (*S80*) with relatively narrow size distributions (see Figure S4). Note that to facilitate the analysis, the density of nanoparticles in the TEM grid is much higher than in the studied granular films. The size is somewhat approximate since the nanoparticles for TEM analysis are unprotected and hence they tend to oxidize. The Co nanoparticle size was estimated assuming complete oxidation of the particles in the TEM grid to CoO and using Co and CoO bulk density values.

*Magnetic characterization.* The magnetic measurements were carried out using a SQUID magnetometer. Field-cooled (FC) and zero-field-cooled (ZFC) magnetization curves were recorded under $\mu_0 H = 20$ mT upon heating from 10 to 390 K. Magnetic hysteresis loops were measured at different temperatures after cooling down from 390 K to 10 K in a field of $\mu_0 H_{FC} = 5$ T. Note that a linear diamagnetic background arising from the Si substrate is subtracted from the hysteresis loop data.

$H_C$ and $H_E$ were obtained from the extrapolation of the M(H) values close to M = 0. This approach leads to rather small errors (less than 1 mT).

**Mean Field Model**

For the calculation of the temperature dependence of the bulk and surface magnetizations, a simple molecular field model is considered, in which the atoms are distributed in successive spherical shells. Let $A_i$ be the surface area of the considered i$^{th}$ shell, $A_{i-1}$ and $A_{i+1}$ the surface area of the shells preceding and following shell *i*. An atom is assumed to have 12 nearest neighbours in total, $z_i = 6$ in the shell it belongs to, $z_{i-1} = 3\ A_{i-1}/A_i$ in the preceding shell and $z_{i+1} = 3\ A_{i+1}/A_i$ in the following shell. In shell, $i$, the fraction of Co atoms is $\alpha_i^{Co}$ and the fraction of Ni atoms is $\alpha_i^{Ni}$. Similarly, the fraction of Co and Ni atoms in shell $i$-$1$ and $i$+$1$ are $\alpha_{i-1}^{Co}$, $\alpha_{i+1}^{Co}$, $\alpha_{i-1}^{Ni}$, and $\alpha_{i+1}^{Ni}$. At the inner shell, at the interface with the core, characterized by $i = 0$, the exchange coupling with the Co FM moment is neglected and thus $z_{-1} = 0$. The diameter of the inner shell is taken as 5 nm and the distance between successive layers as 0.25 nm. The two most inner layers are assumed to be composed of Co atoms only ($\alpha_i^{Co} = 1\ and\ \alpha_i^{Ni} = 0$), then 3 layers are assumed in which Co and Ni atoms are intermixed (with the $\alpha_i^{Ni}/\alpha_i^{Co}$ ratio successively increasing form 0.25 for i = 3 to 0.75 for i = 5), these are followed by 5 additional layers of pure Ni ($\alpha_i^{Co} = 0\ and\ \alpha_i^{Ni} = 1$).

The molecular field on a Co atom in shell $i$ is equal to:

$$B_i^{Co} = \mu_0 N[z_{i-1}\alpha_{i-1}^{Co}n_{CoCo}<\mu_{i-1}^{Co}> + z_{i-1}\alpha_{i-1}^{Ni}n_{CoNi}<\mu_{i-1}^{Ni}> + z_i\alpha_i^{Co}n_{CoCo}<\mu_i^{Co}>$$
$$+ z_i\alpha_i^{Ni}n_{CoNi}<\mu_i^{Ni}> + z_{i+1}\alpha_{i+1}^{Co}n_{CoCo}<\mu_{i+1}^{Co}> + z_{i+1}\alpha_{i+1}^{Ni}n_{CoNi}<\mu_{i+1}^{Ni}>].$$

(S1)

where $N$ is the number of Co or Ni atoms per unit volume, equal to 51.7 $10^{27}$ m$^{-3}$. Similar expressions are derived for the molecular field on atoms in the other shells, and for the Ni atoms in the various shells. The molecular field coefficients $n_{CoCo} = 203$ and $n_{NiNi} = 540$ are derived from the values of the Néel temperatures in these compounds ($T_N = 293$ K in CoO and 524 K in NiO). The coefficient of the intermixed layers, $n_{CoNi}$, is taken as $\sqrt{n_{CoCo}n_{NiNi}}$ = 331, where $S^{Co} = \frac{3}{2}$ and $S^{Ni} = 1$ are assumed.

The Co moment in shell $i$ obeys also the classical molecular field expression:

$$<\mu_i^{Co}> = \mathcal{B}_J(x_i) \qquad (S2)$$

where $\mathcal{B}_J(x_i)$ is the Brillouin function, with $J = S^{Co} = \frac{3}{2}$ for the Co atoms in CoO, and $x_i = 2\mu_B S^{Co} B_i^{Co}/k_B T$. Similar expressions hold for the moments in the other shells and for the Ni moments.

The two sets of equations, (S1) and (S2), are solved self-consistently giving the values of the Co and Ni magnetic moments in each individual layer up to the surface of the FM nanoparticles. The Co magnetization in the bulk, at the surface of CoO and at the surface of CoO-NiO are plotted in Figure S2. Note that no fitting parameters are involved in this calculation.

**Estimation of the CoO shell effective thickness from the fit of $H_E(T)$**

Since the CoO shell is due to the surface oxidation of the Co nanoparticle, it is not expected to be a single crystal, but to be formed by several nanograins (see Fig. S5). TEM observations indicate that the oxide grains at the surface of Co nanoparticles are rather small, 2-3 nm in lateral size. Thus, for simplicity, the oxide grains are taken as small cylinders with their axis perpendicular to the Co nanoparticle surface and 2.5 nm in diameter. Assuming that the oxide grains are not exchange coupled to each other [7], the thermal stability will be linked to the stability of the individual grains rather than to the whole oxide shell volume at the surface of the Co nanoparticles. Consequently, the volume extracted from the fit to the model, $V_{CoO} = 1.25$ nm$^3$, refers to the CoO volume in each individual mixed oxide nanograin, rather than in the whole shell. Thus, a volume $V_{CoO} = 1.25$ nm$^3$ for a cylindrical nanograin 2.5 nm in diameter corresponds to a CoO thickness of 0.25 nm.

**Hysteresis loop for sample R80 at room temperature**

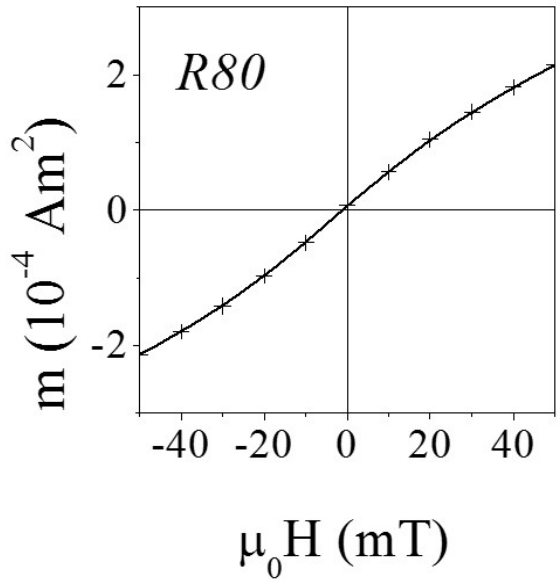

**Figure S1**. Room temperature hysteresis loop for sample R80.

**Mean field calculation of the staggered magnetization**

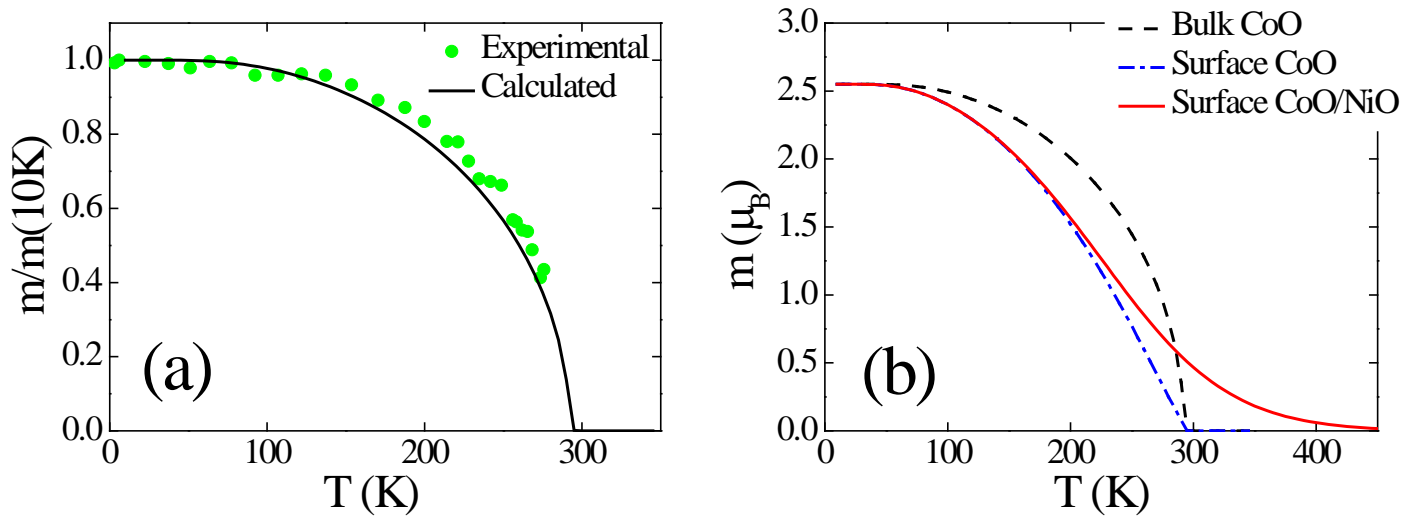

**Figure S2.** (a) Experimental (●) [8] and calculated (—) normalized temperature dependence of the Co staggered magnetization in bulk CoO (b) Temperature dependence of the Co magnetization in CoO: (-----) bulk CoO, (— · —) Co surface atoms in CoO, and (———) Co surface atoms in CoO/NiO.

**Theoretical calculations for the Co/Co-CoO sample**

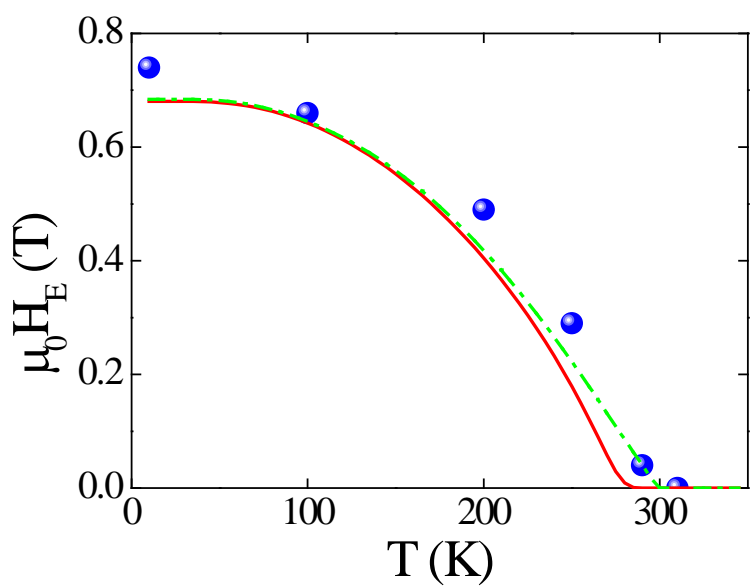

**Figure S3**. Temperature dependence of $H_E$ for the Co/CoO-CoO system: Experimental data (●) [9]; Calculated temperature dependence of $H_E$, neglecting thermal activation (— · —) and taking thermal activation into account (——).

**Morphological characterization**

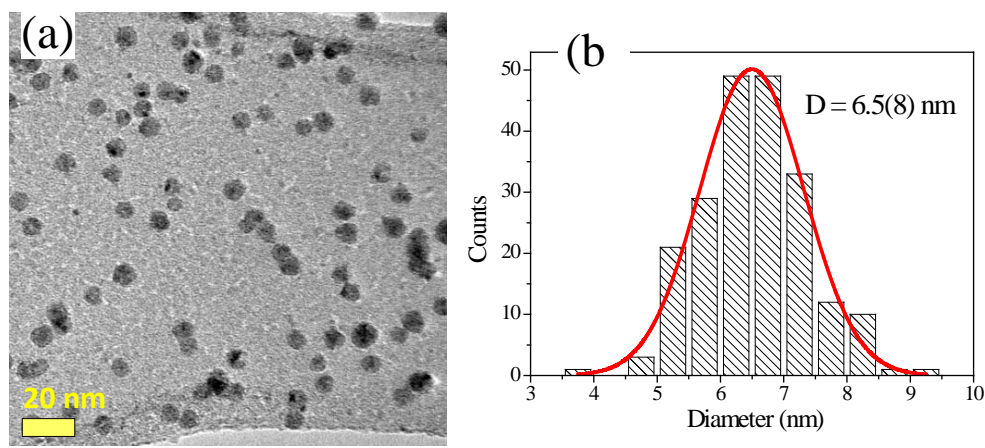

**Figure S4**. Transmission electron micrograph (a) and particle size distribution histogram (b), of the *S50He* cobalt nanoparticles. Note that these particles are partially oxidized (after exposure to ambient conditions), consequently the particles appear larger than they are.

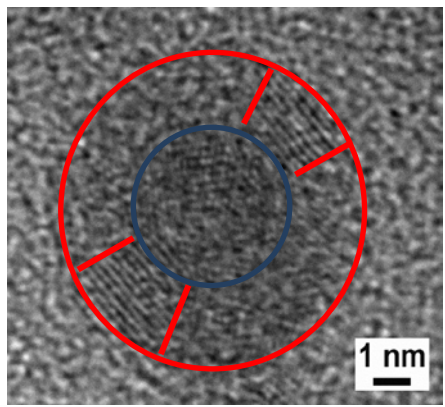

**Figure S5**. High-resolution transmission electron micrograph of a single Co/CoO nanoparticle. The Co core is shown in blue and two of the CoO grains in the shell are highlighted in red.